\def\mearth{{\rm\,M_\oplus}}
\def\rearth{{\rm\,R_\oplus}}

\def\lsun{{\rm\,L_\odot}}
\def\gsim{~\rlap{$>$}{\lower 1.0ex\hbox{$\sim$}}}
\def\lsim{~\rlap{$<$}{\lower 1.0ex\hbox{$\sim$}}}

\def\eg{{\it e.g.\ }}

\def\ie{{\it i.e.\ }}

\documentclass{emulateapj}
\usepackage{graphicx}
\usepackage{amsmath}
\begin{document}

\title{Comparative Habitability of Transiting Exoplanets}
\author{Rory Barnes\altaffilmark{1,2,3}, Victoria S. Meadows\altaffilmark{1,2}, Nicole Evans\altaffilmark{1,2}}
\altaffiltext{1}{Astronomy Department, University of Washington, Box 951580, Seattle, WA 98195}
\altaffiltext{2}{NASA Astrobiology Institute -- Virtual Planetary
Laboratory Lead Team, USA}
\altaffiltext{3}{E-mail: rory@astro.washington.edu}

\begin{abstract}

Exoplanet habitability is traditionally assessed by comparing a
planet's semi-major axis to the location of its host star's
``habitable zone,'' the shell around a star for which Earth-like
planets can possess liquid surface water. The {\it Kepler} space
telescope has discovered numerous planet
candidates near the habitable zone, and many more are expected from missions such as {\it K2}, {\it TESS} and {\it PLATO}. These candidates often require significant follow-up
observations for validation, so prioritizing planets for
habitability from transit data has become an important aspect of the
search for life in the universe. We propose a method to compare transiting planets for their potential
to support life based on transit
data, stellar properties and previously reported limits on planetary
emitted flux. For a planet in radiative equilibrium, the emitted flux
increases with eccentricity, but decreases with albedo. As these
parameters are often unconstrained, there is an ``eccentricity-albedo
degeneracy'' for the habitability of transiting exoplanets. Our method
mitigates this degeneracy, includes a penalty for large-radius
planets, uses terrestrial mass-radius relationships, and, when
available, constraints on eccentricity to compute a number we call the
``habitability index for transiting exoplanets'' that
represents the relative probability that an exoplanet could support liquid surface water. We
calculate it for {\it Kepler} Objects of Interest and find that
planets that receive between 60--90\% of the Earth's incident
radiation, assuming circular orbits, are most likely to be
habitable. Finally, we make predictions for the upcoming {\it TESS} and
{\it JWST} missions.

\end{abstract}

\section{Introduction}
\label{sec:intro}

The discovery of an inhabited exoplanet is a major goal of modern
astronomy and astrobiology. To that end, NASA and ESA have
developed near-term strategies to discover terrestrial planets
in the habitable zones (HZs) of their parent stars with spacecraft
such as {\it Kepler}, {\it K2}, {\it TESS} and {\it PLATO}, and
to spectroscopically measure the atmospheric properties of transiting
exoplanets with {\it JWST}. Meanwhile, ground-based surveys like
MEarth could discover $\sim 1$ additional planet from the ground
\citep{Berta13}. The {\it Kepler} spacecraft has discovered a bounty
of exoplanets, perhaps over 4000, orbiting primarily FGK stars
\citep{Batalha13}, although only a few are confirmed and potentially habitable
\citep[\eg][]{Borucki13,Torres15}. These results, as well
as those from radial velocity (RV) surveys by the HARPS and
HARPS-N instruments \citep[\eg][]{Bonfils13}, lend confidence to the
notion that {\it TESS} will indeed discover tens of potentially
habitable super-Earths (1--2~$\rearth$~and rocky) in the HZs of nearby,
bright stars \citep{Ricker14,Sullivan15} that will be amenable for
{\it JWST} follow-up
\citep{Deming09,Misra14Dimers,Cowan15}.

Absorption features in the transmission spectrum of a habitable
Earth-like planet are easiest to detect if the host star is apparently
bright \citep{Deming09}. Unfortunately, the vast majority of {\it
Kepler} Objects of Interest (KOIs) orbit faint host stars hundreds of
parsecs from Earth, and despite significant effort to validate
potentially habitable planets
\citep[\eg][]{Borucki12,Borucki13,Quintana14,Torres15,Barclay15,Jenkins15}, {\it
Kepler} has probably not discovered a viable target for transmission
spectroscopy. Thus, the {\it K2} and {\it TESS} missions are more
likely to discover appropriate targets.  As Earth-sized planets have transit
depths near the detection limit, we should expect many false
positives, and the full vetting process for these candidates could
require a massive effort.

A transit signal's planetary origin requires validation by other
means, such as RV measurements
\citep{Batalha11,Howard14,Pepe14}, photometry by other spacecraft
\citep{Ballard11}, adaptive optics
\citep[\eg][]{Kraus14}, etc. If we are to achieve the goal of
identifying molecular gases in the transmission spectrum of a
potentially habitable Earth-sized exoplanet, these follow-up
observations must be made. As these measurements can be expensive,
challenging and time consuming, prioritizing targets can increase the
probability of success in the search for life in the universe. The
goal of this paper is to lay out a simple approach to comparative
habitability assessments using basic transit and stellar data to
compare transiting planets in terms of their potential to
support life. We define a ``potentially habitable'' planet to be one
that is mostly rock, with a small ($\lesssim~100$~bar), high
molecular weight atmosphere, and with energy sources and an internal structure 
such that the surface temperature and pressure permit liquid water for
geological timescales.

Both {\it K2} and {\it TESS} should discover $\sim 10$ rocky planets
in the HZ that could be observed with {\it JWST}
\citep{Ricker14,Beichman14,Sullivan15}. The {\it K2} mission can discover
planets in the HZ of their parent star, but they are likely to be
relatively large and orbiting relatively dim M dwarf stars
\citep{Beichman14}. {\it TESS} will find smaller planets around
brighter stars and has been designed to search for planets that will
be easiest to observe with {\it JWST}, see $\S$~\ref{sec:future}. In
both cases (and for {\it Kepler}), transit data combined with stellar
properties provide relatively limited information for assessing
habitability. We expect transit data to accurately constrain the
orbital period $P$, transit depth $d$, and transit duration $D$, and
for the host star's surface gravity log($g$), radius $R_*$, and
effective temperature $T_*$ to be known. From these 6 parameters we
must prioritize targets for their astrobiological interest.

In some cases, more information will be available. The high cadence
{\it TESS} data should also provide the impact parameter $b$. This
addition allows an estimation of the minimum eccentricity $e_{min}$ of
the orbit \citep{Barnes07,Barnes15_tides}. Other transiting exoplanets may
also be detected in some systems, and enforcing stability in those
systems can constrain the maximum eccentricity $e_{max}$
\citep[\eg][]{BarnesQuinn01,Shields15}. When available these data should also be
leveraged in any prioritization scheme.

Having identified the available parameters, the next step is to
determine how to prioritize planets in terms of habitability.
Traditionally, exoplanet habitability is
assessed by comparing a planet's semi-major axis $a$ and host star
luminosity $L_*$ to the location of the HZ, a shell around a star in
which an Earth-like planet could support liquid water
\citep{Kasting93,Selsis07,Kopparapu13,Kopparapu14}. Habitability
assessment using the HZ tends to be binary: Planets in the HZ are
potentially habitable; those outside it are not. Thus, the HZ itself
does not provide the opportunity to determine which planets inside are
most likely to support life, nor does it explicitly tie transit observables to assessment of a planet's potential habitability.

To compare planets' potentials for habitability, one must
quantify the probability that the appropriate conditions are
satisfied. This likelihood is extremely complicated as planetary
surfaces evolve chaotically over timescales ranging from seconds to
Gyr, and distance scales ranging from the atomic to the
galactic. Thus, a proper quantification that folds in all relevant
aspects of composition, evolution, and environment is still very
challenging \cite[see \eg][]{HornerJones10}. However, we can use
previous results to build a simple assessment scheme that depends on
transit observables, stellar properties, and the probability that a
planet is terrestrial.

If a self-consistent terrestrial planet model shows that an
observationally-allowed combination of stellar luminosity, orbital
properties, and albedo can permit surface water, then
by definition that planet is potentially habitable. In practice,
Earth-like planets are generally used to explore the limits of habitability, but other assumptions
have been made, such as slow rotation and a dry surface
\citep{Joshi97,Abe11,Yang13}. Traditionally, the HZ is computed
with 1-D climate-photochemical models
\citep{Kasting93,Selsis07,Kopparapu13}, which have found that both HZ
limits can be quantified in terms of radiation flux. In
particular, the outgoing infrared radiation flux from the planet $F$,
as a function of planetary mass and radius, has been shown to bound
the HZ \citep{Kopparapu14}. The
value of $F$ can be estimated from the transit and stellar data, and
thus we may use astronomical measurements of the star and the
planetary orbit to quantify the incident radiation, and then
estimate the probability that observational data are consistent with
flux limits from models of habitable planets.

The value of $F$ depends on $L_*$, $a$, the albedo $A$ and the
eccentricity $e$:
\begin{equation}
F = \frac{L_*(1-A)}{16\pi a^2\sqrt{1-e^2}},
\label{eq:surfflux}
\end{equation}
where we have assumed that the emitted flux is equal to one-quarter
the absorbed flux (the ratio of the cross-sectional area that absorbed
the flux to the area of the emitting surface) and have averaged over
the orbital period \citep{Berger93}, \ie the planet quickly advects
absorbed radiation to the anti-stellar hemisphere and is in radiative
equilibrium. Note how $F$ depends on $e$ and $A$: Higher eccentricity
will lead to higher fluxes, while higher albedo will lead to lower
fluxes. As neither parameter can be well-constrained by transit data,
there exists an ``eccentricity-albedo degeneracy'' for potentially
habitable transiting exoplanets. We can mitigate this degeneracy by
employing constraints on $e$, but $A$ is usually unconstrained by the
observational data.

Transit data alone do not provide a constraint on planetary density $\rho$, and hence without additional information we do not know if an
exoplanet is rocky or gaseous.  As terrestrial
planets are the more likely abodes of life, we will give lower
priority to larger planets that are more likely to be
``mini-Neptunes,'' \ie small planets dominated by gases and ices
\citep{Barnes09}. Some data and analyses are starting to point to a transition in
the range of 1.5--2~$\rearth$
\citep{Marcy14,LopezFortney14,WeissMarcy14,Rogers15}, but the probability that a planet is rocky
$p_{rocky}$ cannot be well-constrained by transit data alone at this time.

We assume that for a planet to possess surface habitability, then it
must 1) emit between the limiting fluxes of the HZ, and 2) be
terrestrial. We then wish to calculate the likelihood that these
two conditions are met based on the available data. We encapsulate
this likelihood in a parameter we call the ``habitability index for
transiting exoplanets'' (HITE).

In addition to a planet being theoretically able to support life,
prioritization should also fold in detectability. As we show below,
many KOIs have large HITE values but their host stars are likely too
faint for {\it JWST} or any other approved telescope. In
$\S$~\ref{sec:future} we focus on the design constraints of {\it K2},
{\it TESS}, and {\it JWST} to evaluate the likely apparent properties
of a viable target for the latter mission.

This paper is organized as follows. In $\S$~\ref{sec:habidx} we define
the HITE. In $\S$~\ref{sec:KOI} we calculate its value for some {\it
Kepler} Objects of Interest (KOIs). In $\S$~\ref{sec:future} we
consider our results in tandem with the {\it JWST} design, and
evaluate the prospects for {\it TESS} planets. In
$\S$~\ref{sec:discussion} we discuss the method and its
implications, and in $\S$~\ref{sec:concl} we draw our conclusions.

\section{The Habitability Index for Transiting Exoplanets}
\label{sec:habidx}

In this section we describe how to transform transit data into
parameters relevant for planetary habitability and then into the
HITE. We assume that in every case the values of $P$, $d$, $D$, 
log($g$), $R_*$ and $T_*$ are known. Our approach is to 1) apply all
transit and stellar data to calculate quantities such as $L_*$ and
planet radius $R_p$, 2) estimate planetary masses $M_p$ from scaling
laws, 3) identify any constraints on $e$, 4) calculate $F$ over the
permissible range of $e$ and $A$ values and determine the fraction of
the total parameter space with habitable fluxes, and 5) penalize large
planets that may be mini-Neptunes. The final step is to use these
intermediate results to calculate the overall probability $H$ that the
planet has an emitted flux in the habitable range and is terrestrial.
In essence, we are performing the transformations $(P,d,D,R_*,$
log($g$)$,T_*)~\rightarrow~(L_*,a,e,A,R_p)~\rightarrow~H$.
The following subsections describe how we calculate the two
habitability flux limits $F_{min}$ and $F_{max}$, $A$, $e$, $R_p$, $M_p$,
$p_{rocky}$, and $H$.

\subsection{The Limiting Fluxes}

The value of $F_{max}$ is set by a process called the runaway
greenhouse. A planetary atmosphere in a runaway greenhouse has a water
vapor saturated atmosphere that is transparent in the visible, but
optically thick in the infrared, which allows stellar energy to reach
the solid surface, but the thermal photons emitted by the surface
cannot radiate directly back to space. These trapped photons heat the
surface until it reaches a temperature that is high enough for its
blackbody emission to peak in the near-infrared where the density of
ro-vibrational water bands diminishes enough that radiation can escape
to space. However, for this new equilibrium to be achieved, the
surface temperature must reach $\sim 1500$~K, and hence the planet is
no longer habitable
\citep{Simpson27,Nakajima92,Kasting93,Abe93,Abe11,Kopparapu13,Goldblatt15}.

For this study, we will use the analytic runaway greenhouse flux limit
proposed by \cite{Pierrehumbert10}:
\begin{equation}
F_{max} = B\sigma\Big(\frac{l}{2R\ln(P_*\sqrt{\kappa}{P_0g})}\Big)^4,
\label{eq:rgflux}
\end{equation}
where $B$ is a coefficient of order unity that forces the analytic expression to match detailed radiative transfer models, $\sigma$ is the Stefan-Boltzmann
constant, $l$ is the latent heat capacity of water, $R$ is the
universal gas constant, $P_0$ is the pressure at which the water vapor
line strengths are evaluated, $g$ is the
acceleration due to gravity at the surface, and $\kappa$ is a gray
absorption coefficient. $P_*$ is a scaled pressure given by
\begin{equation}
P_* = P_{ref}e^{\frac{l}{RT_{ref}}},
\label{eq:scaledpressure}
\end{equation}
where $P_{ref}$ and $T_{ref}$ correspond to a (pressure, temperature) point on the saturation vapor curve for water. In accordance with \cite{Pierrehumbert10} we set $B =
0.7344$, $\kappa = 0.055$, $T_{ref} = 273.13$~K and $P_{ref} =
610.616$~Pa. Note that $F_{max}$ is an implicit function of $R_p$ and
$M_p$ through $g$. We chose this model because it is
analytic, can be quickly calculated, and gives similar
values to more complicated models.

\begin{figure}[h]
\centering
\begin{minipage}{2.8in}
\resizebox{2.8in}{!}{\includegraphics{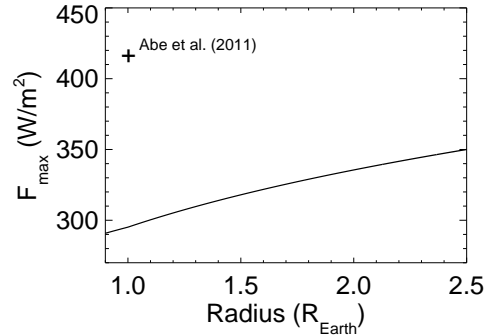}} 
\end{minipage}
\caption{Critical flux to trigger a runaway greenhouse, $F_{max}$, as a function of planetary radius and assuming Earth-like composition. The solid curve is the value predicted by \cite{Pierrehumbert10} and \cite{Sotin07}, while the cross marks the value found via 3-D global circulation models by \cite{Abe11} that is applicable to dry Earth-sized planets. Planets below the curve or cross are potentially habitable.} 
\label{fig:RG}
\end{figure}

Transit and stellar data provide $R_p$, but not $M_p$, so we
must determine {\it it} by other means. Many terrestrial planet
mass-radius relationships are available (see \cite{Barnes13} for a
review), and we use the Earth-like compositional model of
\cite{Sotin07}, in which case
\begin{equation}
\label{eq:massrad}
\frac{M_p}{\mearth} = \left\{ \begin{array}{rl}
 \Big(\frac{R_p}{\rearth}\Big)^{3.268}, &\mbox{ $R_p \le 1~\rearth$,} \\
 \Big(\frac{R_p}{\rearth}\Big)^{3.65}, &\mbox{ $R_p > 1~\rearth$.}
       \end{array} \right.
\end{equation}
With these assumptions and the available stellar and transit data, we
are able to compute $F_{max}$. Note that for this mass-radius model,
$F_{max}$ increases with radius, see Fig.~\ref{fig:RG}.

An alternative runaway greenhouse limit of 415~W/m$^2$ was proposed
for ``dry'' planets by \cite{Abe11}. They only considered an Earth
mass and radius planet and used 3-D global circulation models to
calculate $F_{max}$. As they did not consider a range of masses and
radii, this model cannot be applied to an arbitrary exoplanet. We will
show in $\S$~\ref{sec:discussion} that our prioritization scheme is
largely independent of the choice of dry or wet exoplanets.

At the outer edge of the HZ, the available energy can be too small to
avoid planet-wide glaciation
\citep{Kasting93,Shields14}. \cite{Kopparapu14} examined the outer HZ
as a function of $M_p$ and found $F_{min}~\approx~67$~W/m$^2$ for $0.1
\le M_p \le 5~\mearth$, the latter corresponding to a $1.6~\rearth$
planet. We therefore set $F_{min} = 67$~W/m$^2$ for all terrestrial
planets. \cite{Abe11} found that dry planets could be habitable at
larger stellar distances, but they did not report the minimum outgoing
flux that permitted habitable surfaces.

\subsection{The Minimum Eccentricity}

In some cases, the transit duration and orbital period can be used to
constrain the eccentricity. For well-sampled, high signal-to-noise transits, \eg bright stars with short--cadence photometry, the impact parameter $b$ can be
measured. Combined with $R_*$ and $a$, $e_{min}$ can be calculated
\citep{Barnes07}. This minimum value can be derived by comparing $D$
to the duration predicted if the planet were on a circular orbit,
$D_c$. If $D \ne D_c$, then the planet must be on an eccentric orbit
with transits occurring at an orbital phase in which the instantaneous
azimuthal velocity is not equal to the circular velocity. The
difference between $D$ and $D_c$ is maximized at periastron and
apoastron, and thus the orbit must at least be eccentric enough to
permit the observed duration if the transit longitude
aligns with the orbit's major axis.

The transit duration for a circular orbit is
\begin{equation}\label{eq:durcirc}
D_c = \frac{\sqrt{(R_*+R_p)^2 - b^2}}{\pi a}P,
\end{equation}
where $R_*$ is the stellar radius and $b$ is the impact
parameter. \cite{Barnes15_tides} introduced a convenient parameter $\Delta$, the ``transit duration anomaly'' \citep[see
also][]{Plavchan14},
\begin{equation}\label{eq:Delta}
\Delta \equiv D/D_c,
\end{equation}
which can be combined with Kepler's Second Law to find
\begin{equation}\label{eq:emin}
e_{min} = \left|\frac{\Delta^2 - 1}{\Delta^2+1}\right|.
\end{equation}
This minimum eccentricity can then be used to constrain the possible
range of orbital eccentricities, and hence potential habitability.

The above condition assumes $b$ is well-known, which requires
well-sampled transit ingress and egress and good constraints on limb
darkening. This situation is not met on a per-transit basis for
long-cadence {\it Kepler} and {\it K2} data, however short cadence
{\it Kepler} data and {\it TESS} data may be able to resolve $b$
and hence $e_{min}$ \citep{VanEylenAlbrecht15}.

If $b$ is not well-known, sometimes a minimum eccentricity can still
be calculated. If the transit duration is longer than that predicted
for a circular orbit and a central transit ($b = 0$), then the orbit
must also be eccentric
\citep{Ford08}. As transit observations are biased toward orbital
phases near periastron, this constraint is disfavored, but it can
still be used to constrain the eccentricity distribution of exoplanets
\citep{Moorhead11,Plavchan14}. An analogous derivation to that in Eqs.~(\ref{eq:durcirc}--\ref{eq:emin})
can easily be made and applied when appropriate.

\subsection{The Maximum Eccentricity}

In multiplanet systems, large eccentricities can lead to dynamical
instabilities that destroy the planetary system. This fact can be
exploited to constrain the maximum eccentricity in those systems. In
general, one should perform N-body simulations to prove a certain
orbital architecture is dynamically stable, but that is prohibitively
time-consuming for a large suite of simulations. A faster, but
approximate, approach is to use the Hill stability criterion
\citep{MarchalBozis82,Gladman93,BarnesGreenberg06,BarnesGreenberg07}. This
methodology is only strictly applicable to two-planet systems that are
not in resonance, but it can reasonably approximate stability in more
populated systems \citep{Chambers96}. 

We employ the formalism of \cite{Gladman93} in which Hill stability requires
\begin{equation}
\label{eq:elements}
\zeta^{-3}\Bigg(\mu_1 + \frac{\mu_2}{\lambda^2}\Bigg)(\mu_1\gamma_1 + \mu_2\gamma_2\lambda)^2 > 1 + 3^{4/3}\frac{\mu_1\mu_2}{\zeta^{4/3}},
\end{equation}
where
\begin{equation}
\label{eq:mu}
\mu_i = \frac{m_i}{M_*},
\end{equation}
\begin{equation}
\label{eq:alpha}
\zeta = \mu_1 + \mu_2,
\end{equation}
\begin{equation}
\label{eq:gamma}
\gamma_i = \sqrt{1 - e_i^2},
\end{equation}
and
\begin{equation}
\label{eq:lambda}
\lambda = \sqrt{\frac{a_{out}}{a_{in}}}.
\end{equation}
The subscript $i$ denotes a planet, $M_*$ is the stellar mass, $a_{in}$ and $a_{out}$ are the
semi-major axes of the inner and outer planet of a given pair,
respectively. We can rearrange Eq.~(\ref{eq:elements}) to determine
the maximum eccentricity,
\begin{equation}
\label{eq:emax}
e_{max} = \Bigg(1 - \Bigg(\frac{\Big(\frac{1+3^{4/3}\big(\frac{\mu_1\mu_2}{\zeta^{4/3}}\big)}{\zeta^{-3}\big(\mu_1+\frac{\mu_2}{\lambda^2}\big)}\Big)^{1/2} - \mu_2\gamma_2\lambda}{\mu_1}\Bigg)^2 \Bigg)^{1/2}.
\end{equation}
When applicable, we apply this condition to our estimates of
likelihood of habitability{\footnote{Code to calculate Hill stability
boundaries is publicly available at
https://github.com/RoryBarnes.}. For planets with $R_p < 2.5~\rearth$,
we use Eq.~(\ref{eq:massrad}) to calculate the mass, while for larger
planets we assume a density of 1 g/cm$^3$ \cite[see
\eg][]{Lissauer11}. If a planet has both interior and exterior
planets, we calculate $e_{max}$ for both and use the smaller.

\subsection{The Eccentricity Distribution}

The eccentricity distribution of known exoplanets is not flat, and hence we
should not treat all permissible eccentricity values equally. Tidal
circularization impacts planets in the HZ if $M_* \lesssim
0.2$~M$_\odot$~\citep{Barnes08,Barnes13}, where the orbital periods are $\lesssim
15$~days. The eccentricity distribution of known exoplanets with
orbital periods longer than 15 days, as of 4 June 2013\footnote{Data
from http://exoplanets.org} is peaked near zero with a long tail to
nearly 1, as shown by the histogram in Fig.~\ref{fig:probecc}. We
weight eccentricities by the frequency that that eccentricity is
observed, $p(e)$. We fit the observed data to a third order polynomial
using a Levenberg-Marquardt minimization scheme and find that
\begin{equation}
\label{eq:probecc}
p(e) = 0.1619 - 0.5352e + 0.6358e^2 - 0.2557e^3.
\end{equation}
This fit has a $\chi$-squared of 1.344, and is shown by the dashed
line in Fig.~\ref{fig:probecc}. 

\begin{figure}[h]
\centering
\begin{minipage}{2.8in}
\resizebox{2.8in}{!}{\includegraphics{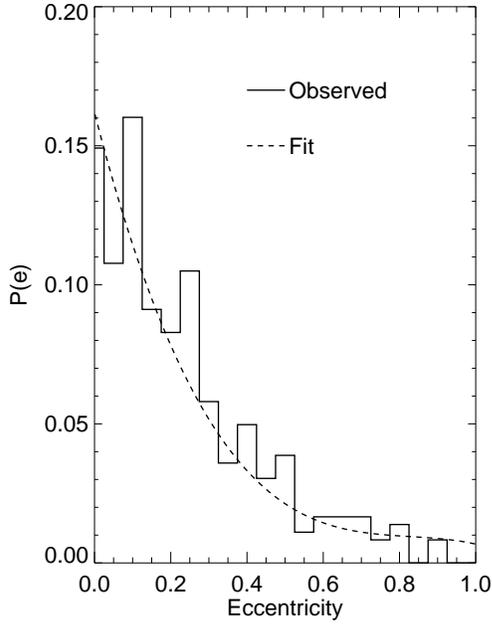}} 
\end{minipage}
\caption{Eccentricity distribution of known exoplanets that are beyond the reach of tidal circularization. The histogram is the observed distribution with a binsize of 0.05, dashed the polynomial fit from Eq.~(\ref{eq:probecc}).} 
\label{fig:probecc}
\end{figure}

We stress that this distribution may not represent that of terrestrial
exoplanets. The eccentricities of those worlds are difficult to
measure with RV data, and secondary eclipses are often too shallow to
be measured individually \citep{SheetsDeming14}. Small planets with large companions could possess larger
eccentricities due to planet-planet gravitational interactions. On the other hand, the formation process could
force their eccentricities to be lower. Current understanding does
not permit a detailed exploration of these trade-offs, so until
additional data become available we assume the giant planet
eccentricity distribution is a reasonable approximation for the terrestrial planet distribution.

\subsection{Is the Planet Rocky?}

The next concern is the likelihood that a given planet
possesses a large gaseous envelope. Considerable effort has been
expended to identify the density of small exoplanets
\citep{Marcy14,WeissMarcy14,Rogers15,Barnes15_tides}, but their nature
is still elusive. These planets do not induce large reflex motions in
their host star, nor strong transit timing perturbations on their
sibling planets, hence direct mass measurements are
difficult. Furthermore, planetary radii can only be known as
well as the stellar radii and some discrepancy exists between reported
stellar properties, \cite[see
\eg][]{Everett13,GaidosMann13}. \cite{Rogers15} advocated that planets
are likely to be predominantly rocky for
$R_p~<~1.6~\rearth$. On the other hand, Kepler-10~c has a radius of
$2.35~\rearth$ and a mass of $17~\mearth$  precluding a thin hydrogen atmosphere \citep{Dumusque14}. In light of these uncertainties and small number
statistics, we will invoke the following reasonable, but admittedly
{\it ad hoc}, model for the likelihood of an exoplanet being
non-gaseous:
\begin{equation}
\label{eq:probrocky}
p_{rocky}(R_p) = \left\{ \begin{array}{rl}
 1, &\mbox{$R_p~\le~1.5~\rearth$} \\
 (2.5 - R_p), &\mbox{$2.5~\rearth~>~R_p~>~1.5~\rearth$}\\
 0, & \mbox{$R_p~\ge~2.5~\rearth$.}
       \end{array} \right.
\end{equation}
In other words planets smaller than $1.5~\rearth$ do not possess a significant gaseous envelope, and
that likelihood drops linearly to 0 by $2.5~\rearth$. While some small
exoplanets ($R_p < 1.5~R_\oplus$) may still be gaseous, lowering the
probability from unity does not significantly affect the relative rankings.

\subsection{The Eccentricity-Albedo Degeneracy}

Eq.~(\ref{eq:surfflux}) demonstrates the inherit degeneracy between
eccentricity and albedo when calculating the emitted flux: As $A$
increases, $F$ decreases, but as $e$ increases, $F$ increases. To
address this eccentricity-albedo degeneracy, we calculate $F$ from a
grid of $A$ and allowed $e$ values. Essentially we are calculating $F$
for all plausible ($A,e$) pairs to calculate the fraction of pairs
that are potentially habitable. \cite{WilliamsPollard02} used 3-D global climate model (GCM)
to determine that the Earth could be habitable up to at least $e=0.7$,
and so we explore the large, but plausible, range $0~\le~e~\le~0.8$
(recall from $\S$~\ref{sec:habidx}.4 that large $e$ values are downweighted anyway). The albedo
distribution of habitable planets is completely unknown. The Earth's
bond albedo is near 0.3 and is slightly variable due to cloud and vegetation
coverage
\citep{Palle04}. The Moon has a bond albedo of 0.11; Venus' is $\sim 0.8$ \citep{Tomasko80,Moroz85} and
Titan's is 0.265 \citep{Li11}. Reflective hazes or clouds are possible on
habitable planets, perhaps even during the Archean and Proterozoic eras of Earth's history
\citep{Zerkle12,Arney15,Izon15}. The nature and extrema of albedos is
unknown, and therefore we adopt generous limits of
$0.05~\le~A~\le~0.8$ and assume that all possibilities are equally
likely.

We now combine all these concepts to create an assessment scheme for
potentially habitable planets.  We first define an intermediate
parameter $h$ such that
\begin{equation}
\label{eq:littleh}
h(A,e) = \left\{ \begin{array}{rl}
 1, &\mbox{$F_{min} < F < F_{max}$} \\
 0, &\mbox{\textrm{otherwise,}}
       \end{array} \right.
\end{equation}
and calculate it over $0.05~\le~A~\le~0.8$ and $e_{min}~\le~e~\le~e_{max}$.  The habitability index for transiting exoplanets is the fraction of parameter space
for which $h=1$ times the probability the planet is rocky:
\begin{equation}
\label{eq:hidx}
H = \frac{\sum{h_jp_j(e)}}{\sum{p_j(e)}}p_{rocky},
\end{equation}
where $j$ indexes ($A,e$), and $p_j(e)$ is the eccentricity probability distribution in Eq.~(\ref{eq:probecc}). We sample with a resolution of 0.01 in both
parameters and find that finer sampling does not change the results
significantly. We have made available a website to calculate $H$ for newly-discovered exoplanets\footnote{http://vplapps.astro.washington.edu/hite}.

Eccentricity constraints will not always be available, and so we define
a second $H$ parameter that does not include limits on the range of
$e$ that is scanned. The {\it Kepler} and {\it K2} missions observe
most stars with 30 minute exposures, which prevent tight constraints
on $b$ and hence $e_{min}$. {\it TESS} will take 2 minute exposures
and hence will provide tighter constraints on $e_{min}$. For the
short-term, a version of $H$ without eccentricity constraints is a
better metric, and we will define the parameter $H'$ to be the HITE
without the minimum and maximum eccentricity constraints. We will use $H'$ to examine
KOIs in the next section, but $H$ will ultimately be a better
parameter for performing comparative habitability in the upcoming {\it
TESS} and {\it PLATO} eras.

As an example, consider Kepler-62~f \citep{Borucki13} and KOI-5737.01
\citep{Batalha13}. The former is a $1.43~\rearth$ planet orbiting
0.699~AU from a $0.2~\lsun$ star. The latter is a $1.43~\rearth$
planet orbiting 1.012~AU from a $1.01~\lsun$ star. For both planets
$F_{max}$ = 315~W/m$^{-2}$. The nominal system parameters place both
planets in their HZ, but which is more likely to be habitable?

Figure~\ref{fig:compare} shows the expected values of $F$ over a
range of $A$ and $e$ for Kepler-62~f in black and KOI~5737.01 in
red. The contours show $F$ in W/m$^{-2}$ and the solid contours
represent the only possible limiting flux value for that planet
in this parameter space. Habitable conditions could exist for ($A,e$)
combinations below an $F_{min}$ contour and above an $F_{max}$
contour. Ignoring eccentricity limits, KOI~5737.01 appears to be a
better candidate for habitability as a larger fraction of
($A,e$) combinations permit habitability than for Kepler-62~f. Their
two values for $H'$ are 0.66 and 0.92, for Kepler-62~f and
KOI~5737.01, respectively, and the latter is the higher priority
object.

\begin{figure}[h]
\centering
\begin{minipage}{2.8in}
\resizebox{2.8in}{!}{\includegraphics{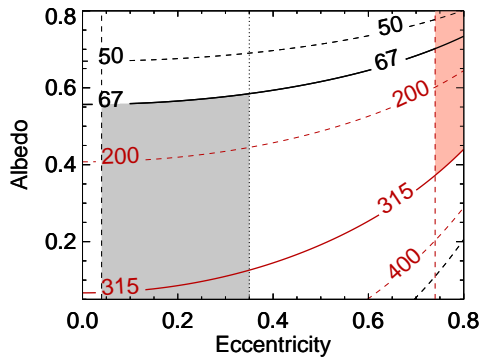}} 
\end{minipage}
\caption{The outgoing flux from two {\it Kepler} targets in terms of eccentricity and albedo. Kepler-62 f in black and KOI 5737.01 in red. The contour lines denote the outgoing flux in W/m$^{-2}$, with the solid contours representing the only limiting flux value to habitability that is appropriate for that planet. The dashed vertical lines are $e_{min}$. The dotted black line is $e_{max}$ for Kepler-62 f. KOI 5737.01's $e_{max}$ is 1. The shading represents the portion of parameter space that is potentially habitable when eccentricity constraints are invoked.}
\label{fig:compare}
\end{figure}

However, the situation is markedly different if the eccentricity
constraints are invoked. The values of $e_{min}$ are 0.04 and 0.74 for
Kepler-62~f and KOI~5737.01, respectively. Orbital stability gives
$e_{max} = 0.35$ for Kepler-62~f, which has been confirmed by N-body
integrations in \cite{Shields15}, but KOI~5737.01 is isolated so its
maximum eccentricity is set only by orbits that intersect the host
star. In Fig.~\ref{fig:compare}, the vertical dashed lines represent
$e_{min}$ and dotted $e_{max}$, and the shaded regions are potentially
habitable. The constraints on $e$ change their likelihoods of
habitability. KOI~5737.01 is habitable throughout most of the full
($A,e$) space, but in the narrow range between $e_{min}$ and 0.8, only
about half of ($A,e$) pairs is habitable. Kepler-62~f's habitable
parameter space is only slightly smaller. Including
$e$ constraints, the values of $H$ are 0.65 and 0.48 for Kepler-62~f
and KOI~5737.01, respectively. Thus, with the eccentricity
constraints, Kepler-62~f is a higher priority object. It should be
noted that the large value of $e_{min}$ for KOI~5737.01 could be
artificially high because long cadence {\it Kepler} data do not
provide robust constraints on impact parameters
\citep{Barnes15_tides}.  Nonetheless, this example is illustrative of
how incorporating all observational constraints could change a
prioritization strategy.

\section{Comparative Habitability of KOIs}
\label{sec:KOI}

In this section we consider the entire {\it Kepler} sample as of 17
Aug 2015, including confirmed and unconfirmed planets, but no false
positives\footnote{http://exoplanetarchive.ipac.caltech.edu.}. If
a KOI has been validated, we use the ``Confirmed Planets'' data
instead of the original KOI data. We cut this sample by requiring the
equilibrium temperature $T_{eq}$, as reported by the {\it Kepler} team
(which assumed $A = 0.3$), to lie between 150 and 400~K and $R_p$ to
be less than 2.5~$\rearth$. These cuts are intended to be generous and
identify all KOIs with a even a small chance of
habitability. These cuts leave 268 KOIs, the 10 with the highest
$H'$ values are shown in Table~1, with the full table available
on-line. 

Figure~\ref{fig:hzkep} shows the conventional approach to identifying
habitable planets: the semi-major axes of the planets are compared to
the star's semi-major axis limits from 1-D photochemical-climate
models. Planets in the green regions are potentially habitable; those
exterior are not. In this case we have used the recent HZ limits
proposed by \cite{Kopparapu13}. This representation is not fully
self-consistent as the HZ limits assumed specific stellar mass-radius
and mass-luminosity relationships, and the inner edge is a function of
planetary radius. Nonetheless, it is clear that the {\it Kepler}
spacecraft has discovered numerous planetary candidates in the
classical HZ. The KOIs that pass our cuts tend to be large and
interior to the HZ. These features are primarily due to the well-known
biases associated with transit detection as well as our liberal upper
bound on albedo. The large number of small planet candidates near the
HZ motivates the creation of a comparison scheme for potential
habitability.

\begin{figure}[h]
\centering
\begin{minipage}{2.8in}
\resizebox{2.8in}{!}{\includegraphics{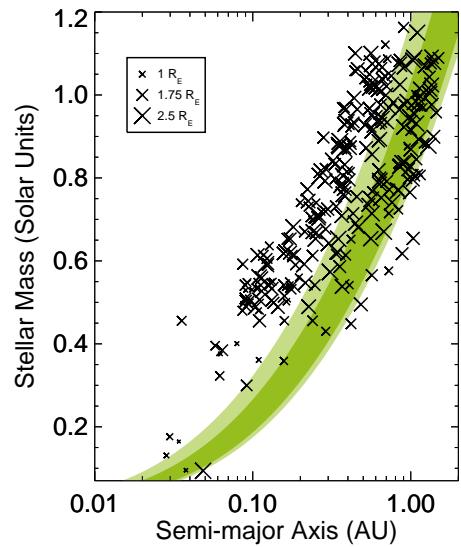}} 
\end{minipage}
\caption{Comparison of potentially habitable KOIs to the HZ. Symbol size is proportional to planetary radius as shown. Dark green is a conservative estimate of the limits of the HZ; light green optimistic \citep[see][]{Kopparapu13}.}
\label{fig:hzkep}
\end{figure}

Of our 268 objects, 194 had $H' \ge 0.01$ and no object
reached $H' = 1$. In Fig.~\ref{fig:habidx} we plot $H'$ against the
incident stellar radiation, assuming the orbit is circular,
$S_{circ}$, and scaled to the Earth's value, $S_\oplus$, for all
objects with $H' > 0.01$. The peak of $H'$ occurs at
$S_{circ}~\sim~0.8$, and objects near the peak should be the highest
priority for follow-up, modulo observational constraints, see
$\S$~\ref{sec:future}.

Figure \ref{fig:habidx} also contains information regarding which
limit is reducing $H'$. For $0.5 \le S_{circ} \le 1$ both flux limits
can be important. In this range $H'$ can nearly reach unity. Objects
with $S_{circ} > 2.5$ have less than a 20\% chance of habitability by
our analysis, and generally require $A~\gsim~0.7$. For individual
systems the uncertainties in $S_{circ}$ can be quite large due to
large uncertainties in stellar parameters \cite[\eg][]{Kane14}. For
the planets we consider here, the average $1\sigma$ uncertainty in
$S_{circ}$ is $\pm~36$\% and so comparative habitability of KOIs is
severely hampered by poor stellar characterization and one should not
place too much trust in individual $H'$ values. Figure
\ref{fig:habidx} primarily shows the trend with $S_{circ}$. Presumably
closer and brighter planet-hosting stars, \eg those discovered by, \eg
{\it TESS}, will have more tightly constrained values of $S_{circ}$.

\begin{figure}[h]
\centering
\begin{minipage}{2.8in}
\resizebox{2.8in}{!}{\includegraphics{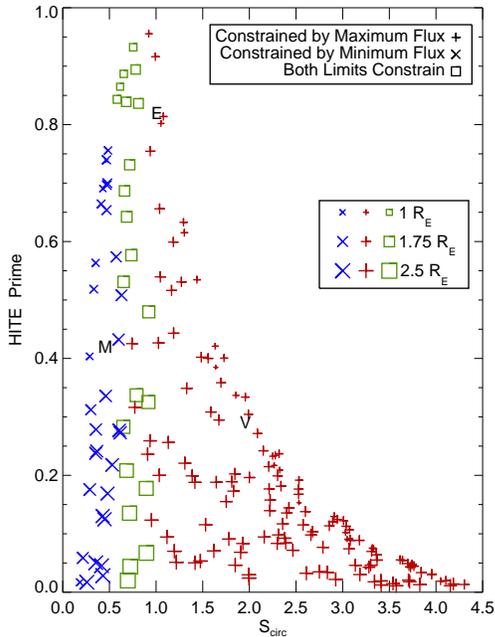}} 
\end{minipage}
\caption{$H'$ values for KOIs as a function of the incident stellar radiation scaled to Earth's value, and assuming a circular orbit ($S_{circ}$). Crosses indicate planets that may be uninhabitable due to too much absorbed energy, x's too little, and open squares indicate both limiting cases are possible. Symbol size is proportional to planetary radius. The locations of Venus, Earth, and Mars are labeled with a V, E and M, respectively.} 
\label{fig:habidx}
\end{figure}

\begin{table*}
\begin{center}Table 1: Observed and Derived Parameters for Potentially Habitable KOIs\\
\footnotesize
\begin{tabular}{ccccccccccccccc}
\hline
ID & log($g$) & $T_{eff}$ & $R_*$ & $P$ & $d$ & $D$ & $b$ & $S_{circ}$ & $H$ & $H'$ & $J$\\
 &  (cm/s$^2$) & (K) & (R$_\odot$) & (d) & (ppm) & (hr) & & ($S_\oplus$) & & & \\
\hline
KOI 3456.02 & 4.37 & 6008 & 1.06 & 486.1270 & 120.6 & 9.495 & 0.362 & 0.92 & 0.805 & 0.955 & 11.83\\
KOI 7235.01 & 4.60 & 5606 & 0.76 & 299.6658 & 222.3 & 20.410 & 0.759 & 0.76 & 0.000 & 0.932 & 13.50\\
KOI 5737.01 & 4.46 & 5916 & 0.96 & 376.2425 & 185.7 & 4.064 & 0.328 & 0.99 & 0.480 & 0.916 & 12.61\\
KOI 2194.03 & 4.51 & 6038 & 0.92 & 445.2291 & 238.1 & 20.620 & 0.266 & 0.78 & 0.890 & 0.894 & 12.70\\
KOI 2626.01 & 4.91 & 3482 & 0.35 & 38.0972 & 1016.8 & 3.377 & 0.548 & 0.65 & 0.913 & 0.887 & 13.45\\
KOI 6108.01 & 4.39 & 5551 & 0.96 & 485.9230 & 122.9 & 3.756 & 0.452 & 0.61 & 0.000 & 0.865 & 10.90\\
KOI 5948.01 & 4.60 & 5776 & 0.76 & 398.5131 & 249.2 & 5.410 & 0.509 & 0.58 & 0.927 & 0.843 & 12.68\\
KOI 6425.01 & 4.43 & 5942 & 0.95 & 521.1054 & 231.9 & 14.540 & 0.822 & 0.68 & 0.888 & 0.839 & 13.01\\
Kepler-442b & 4.67 & 4402 & 0.60 & 112.3053 & 614.1 & 5.621 & 0.220 & 0.81 & 0.838 & 0.836 & 13.23\\
Kepler-296e & 4.83 & 3572 & 0.43 & 34.1420 & 852.2 & 3.619 & 0.180 & 1.08 & 0.850 & 0.814 & 13.39\\
...\\
Kepler-138d & 4.89 & 3841 & 0.44 & 23.0881 & 596.7 & 1.946 & 0.924 & 2.25 & 0.000 & 0.233 & 10.29\\
KOI 5554.01 & 4.17 & 6113 & 1.28 & 362.2220 & 52.5 & 14.290 & 0.566 & 2.26 & 0.216 & 0.217 & 10.15\\
KOI 7587.01 & 4.46 & 5941 & 0.94 & 366.0877 & 494.6 & 11.033 & 0.776 & 1.03 & 0.178 & 0.200 & 10.34\\
\end{tabular}
\end{center}
\end{table*}
\normalsize

The values of $H'$ for Venus, Earth, and Mars are 0.300, 0.829, and
0.422, respectively. Several KOIs have values larger than Earth,
including the confirmed planet Kepler-442 b. For reference, Kepler-62
f has $H' = 0.66$, Kepler-452 b has $H' = 0.60$, Kepler-186 f has $H'
= 0.40$ and Kepler-22 b has $H' = 0.09$. Note in Table 1 that if
eccentricity constraints are included, the rank of KOI~3456.02 drops
significantly.\\\\

\section{Future Prospects}
\label{sec:future}

\subsection{Application to {\it JWST}}

The first telescope that can measure the atmospheric
composition of a terrestrial exoplanet is {\it
JWST}~\citep{Deming09,KalteneggerTraub09,Misra14Dimers}. Many of the
details of {\it JWST}'s capabilities in regard to spectroscopic
measurements of terrestrial exoplanets have been presented elsewhere
\citep{Beichman14,Cowan15}, but have focused on the length of the
observations. In other words, they assumed a target was already in
hand. However, the path to discovering an appropriate target is
decidedly non-trivial. The best targets will have high $H$
values, be visible year-round and have apparent brightnesses just
below the saturation limit.

The first constraint we consider is apparent brightness. More photons
generally means higher signal-to-noise, but below a certain apparent
magnitude, the detector can saturate and invalidate the
data. \cite{Beichman14} find the minimum $J$-band magnitude is
6.9, but note that future modifications could lower it. Additional
detector issues, such as the maximum number of frames that can be
taken sequentially, are of second order, so we will ignore them here
and refer the reader to
\cite{Beichman14}. We include $J$ band magnitude in Table 1.

The second observability constraint is the total in-transit time that will be available over {\it JWST}'s lifetime. The entire sky is not
always accessible, and hence higher priority should go to objects that
maximize this time. The pathological case
is a planet on a 1 year orbit with transits that occur when the system
is occulted by the Sun. More prosaically, a system could be discovered
such that during {\it JWST}'s lifetime the transits just don't occur
when the target is accessible. Thus, priority is a function of the
transit ephemerides, the system's position on the celestial sphere,
and {\it JWST}'s launch date.

Without a firm launch date, it is unknown how many transits will be
visible.  The field of regard, or the portion of the sky that is
accessible to the telescope at any given time, is limited to solar
angles between $\theta_{min} = 85^\circ$ and $\theta_{max} =
135^\circ$, and can be computed as a function of ecliptic
declination. We use the Space Telescope Science Institute's
Astronomer's Proposal Tool to calculate the observability of potential
{\it JWST} targets\footnote{Available at
http://www.stsci.edu/hst/proposing/apt.}.

Only one confirmed planet, Kepler-138 d (KOI-314 c), has a
significant $H'$ value (0.23) and orbits a star with $J < 11$. However,
transit timing variations indicate a mass that is too low for a rocky
composition
\citep{Kipping14}. Three {\it Kepler} planet candidates are potentially habitable, KOI 5554.01 ($H' = 0.22$),
KOI 6108.01 ($H' = 0.87$) and KOI 7587.01 ($H' = 0.20$), and orbit
bright stars ($J < 11$). These planets are also listed in Table 1. KOI 5554.01 lies close to the maximum flux limit, but
perhaps future vetting could improve its rank. We note that the {\it
Kepler} team reports a stellar temperature of 6100~K, a radius of 1.3
R$_\odot$, but a mass of just 0.87 M$_\odot$. The host star has an
apparent $J$ magnitude of 10.1, and is the brightest target in our
list. During the 2020's transits will occur in December and slowly
slide into November at which point it is no longer accessible to {\it
JWST}. Given {\it JWST}'s current configuration we find transits
between 2018 and 2026 should be observable. With a 14.3 hour duration,
it may be possible to acquire $\sim 125$ hours of in-transit
spectroscopy of this planet, if it is validated. While a detailed
study of this particular case is beyond the scope of this study, we
note that at an estimated distance of 175~pc, this amount of time is
unlikely to be sufficient to detect water bands with {\it JWST}
\citep{Deming09}.

KOI 7587.01 has an orbital period of 366 days and
transits will occur in late June in the 2020's. This is close to {\it
JWST} observational windows, but unless modifications to the
satellite's observational constraints occur, this planet is unlikely
to be observable by {\it JWST}. KOI 6108.01 has an
orbital period of 486 days, which is very close to 4/3 of Earth's
orbital period. During the 2020's transits will occur in February,
June and October, but only the February transits occur in {\it JWST}'s
field of regard. We conclude that KOI 5554.01 is the highest priority
target for validation as it might be the only KOI that can be
characterized spectroscopically with {\it JWST}.

The {\it K2} mission has already discovered one potentially habitable
super-Earth, K2-3 d \citep{Crossfield15}. This planet has a radius of
$1.5~\rearth$ and orbits an M0V star with an orbital period of 44.6
days. We find this planet has $H' = 0.505$ and the star has a
$J$ magnitude of 9.41. We include it in the online data of Table
1. Its value of $H'$ places it 38th in terms of potential
habitability, which is in the top quartile. Unfortunately due to its
location on the ecliptic, K2-3 d is only accessible about 28\% of the
year, so although it transits 8--9 times per year, only 2--3 will be
observable.

\subsection{Predictions for {\it TESS}}

The previous subsection highlights the challenges in detecting
potentially habitable planets that are amenable to {\it JWST}
spectroscopy with {\it Kepler} and {\it K2}. The {\it TESS} spacecraft
\citep{Ricker14} has been designed explicitly to overcome these
difficulties and identify habitable planets of nearby, bright
stars. {\it TESS} is an all-sky transit survey with an observational
footprint that mimics that of {\it JWST}, \ie it prioritizes the
ecliptic poles. In this subsection we apply our methodology to the
predicted yield of {\it TESS} planets \citep{Sullivan15} and calculate
the expected properties of its planet candidates.

For a full review of the {\it TESS} mission, consult
\cite{Ricker14}. Here we review the salient points. The {\it TESS}
design will favor the discovery of potentially habitable planets
around early M dwarf stars with HZ orbital periods $\lesssim~20$~days
and with photometric precision of order a few ppm. Most stars will
only be monitored for 45 days, and hence earlier-type stars have HZs
that are too distant for more than 2 transits to be detected, while
stars later than $\approx$M4 are naturally less abundant
\citep{Henry09}\footnote{http://www.recons.org}. Furthermore, the smaller relative size of M dwarfs produces deeper transits for small terrestrial planets.

\begin{figure}[h]
\centering
\begin{minipage}{2.8in}
\resizebox{2.8in}{!}{\includegraphics{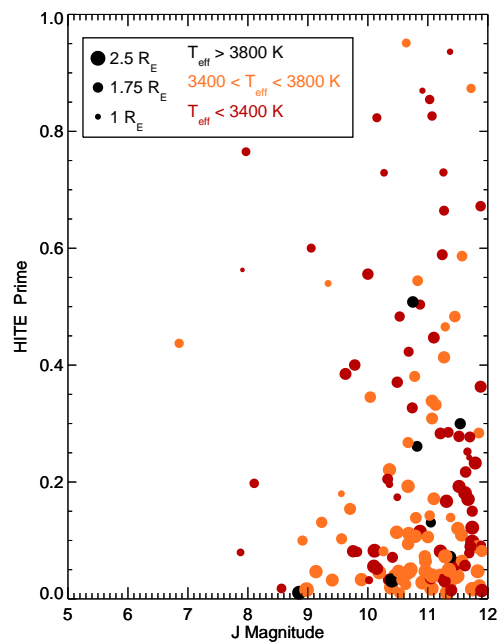}} 
\end{minipage}
\caption{Values of $H'$ for the predicted yield of {\it TESS} planets from \cite{Sullivan15} as a function of $J$ magnitude. Symbol size corresponds to planet size and color to the host star's effective temperature. The colors correspond approximately to K dwarfs and earlier (black), M0--M3 dwarfs (orange), and M4 and later (red).} 
\label{fig:tess}
\end{figure}

Recently, \cite{Sullivan15} conducted an in-depth study to predict the
exoplanet yield from {\it TESS}. They used a galactic model, the
nominal {\it TESS} mission parameters, and planet occurrence rates from
\citep{DressingCharbonneau15}. They found that about 50 planets with
$R < 2~\rearth$ will be discovered with $\sim 5$ in orbit around stars
with $K$-band magnitudes less than 9. 

\cite{Sullivan15} also provided a complete catalog of their predicted yields with enough information to calculate $H'$ (see their Table 6). In Fig.~\ref{fig:tess} we plot the predicted values as a function of $J$. We color code the points by $T_{eff}$ (K dwarfs in black; M0-M3 in orange; M4 and later in red), and the dot size is proportional to planetary radius. Consistent with \cite{Sullivan15}, we find 5 planets with $H' > 0.2$ and the best 2 candidates have a combined value of 1.33, suggesting if both were observed, then probably 1 would be potentially habitable, and its transmission spectrum from {\it JWST} would be very valuable.

\section{Discussion}
\label{sec:discussion}

We have outlined a methodology to prioritize potentially habitable
transiting exoplanets for follow-up with ground- and space-based
telescopes with the explicit goal of identifying potentially habitable
exoplanets amenable to transit spectroscopy. Our
approach relies on published limits to habitability and assumes the
bare minimum of transit and stellar data are available. The 
rate at which the planet's atmosphere can cool to space, which is set by the emitted flux, has been shown to
define the limits of habitability
\citep[\eg][]{Kopparapu14}, and we have shown here how to calculate
$F$ from the basic astronomical data.

To make our habitability comparisons, we have introduced the concept
of a probabilistic $H$ that assigns a number from 0 to 1 to a planet,
with larger numbers indicating a higher chance that liquid surface
water is possible. This approach is markedly different from that of
the classic HZ, which is essentially a binary function: potentially
habitable, or not. A continuous function of habitability will become
more relevant as more candidates become available. This approach is
similar to that in \cite{ShulzeMakuch11}, but our method is tied
directly to the observables and is based on the limits of the
HZ. \cite{Torres15} also considered the likelihood that a planet was
in the HZ and that it is rocky, but they did not consider the
eccentricity-albedo degeneracy, nor did they combine their
probabilities into a single index. Thus, $H$ is the first 
physically-motivated parameter to synthesize observables into a
single number that permits comparative habitability.

We have couched $H$ in terms of the absorbed (and hence emitted)
flux $F$, which is a function of $e$ and $A$. The dependence on $e$ is
weaker than $A$, and $e$ can often be constrained by observations, thus
our results are most sensitive to our choice for the distribution of
$A$. We have treated $A$ as realistically as possible, but should new
insights become available, the concepts presented here should be
revisited. For example, planets with high $A$ may have atmospheric
properties that significantly alter the values of $F_{max}$ and/or
$F_{min}$, \eg $F_{max} = F_{max}(A)$. We encourage future work to
explore the connections between $F_{max}$, $F_{min}$ and $A$.

For known KOIs we examined a wide, but plausible range of ($A,e$)
combinations, ignoring limits on $e$ as the impact
parameters are not well-constrained. Examination of $H'$ and
$S_{circ}$ reveals the best candidates for habitability receive
60--90\% of the Earth's solar constant, which is about the middle of
the classic HZ
\citep{Kopparapu13}. Any small planet in this instellation range
orbiting a bright star should immediately receive intense scrutiny for
validation, stellar characterization, RV follow-up, and
observations that identify the full orbital architecture of the
stellar system.

We also find that KOIs with $S_{circ}~\ge~2.5$ have less than a 20\%
chance of being habitable and those with $S_{circ}~\ge~4$ have less
than a 5\% chance. Previous studies that calculated $\eta_\oplus$, the
average number of terrestrial planets in the HZ per star, have assumed
such planets are habitable \citep[\eg][]{Petigura13}. Our analysis
suggests that these worlds are unlikely to be habitable unless their
albedos are nearly equal to Venus'.

The continuous nature of $H$ can permit a calculation of the
occurrence rate of potentially habitable planets. If we define a
new parameter $\eta_{phab}$ to be the average number of ``potentially
habitable'' planets per star, then its value is the sum of $H$ (or
$H'$) values divided by the total number of stars observed. Current
technology is not sensitive enough to detect all planets of a given
star, but techniques have been developed to remove the observational
biases against small planets with potentially habitable orbits
orbiting FGKM stars
\citep[\eg][]{CatanzariteShao11,Traub12,Petigura13,ForemanMackey14}.
From our study, the sum of all $H'$ values of the KOIs is 49.4,
of which $\sim 10$\% are likely to be false positives
\citep{Fressin13}. Our analysis suggests {\it Kepler} has discovered
$\sim 45$ potentially habitable exoplanets so far. Future
work could remove the observational biases and derive a value for
$\eta_{phab}$.

Our approach can be applied to planets orbiting FGK dwarf stars, and many,
but not late, M dwarfs. Planets orbiting late M dwarfs can have
significantly more complications that impact habitability. While one
can still apply our methodology to those planets, extra phenomena
should be considered if possible. For example, the pre-main sequence
luminosity evolution of M dwarfs can dessicate planetary surfaces
\citep{LugerBarnes15} and extreme tidal heating can trigger a runaway greenhouse \citep{Barnes13} or significantly alter a terrestrial planet's evolution \citep{DriscollBarnes15}. Tidal heating contributes to the energy budget of the atmosphere and would
need to be added to Eq.~(\ref{eq:surfflux}). Expressions for tidal
heating can be found in, {\it e.g.}, \cite{Heller11} and
\cite{Barnes13}. Additionally, tidal circularization sculpts the $e$
distribution in the HZ for stellar masses below $\sim 0.15$~M$_\odot$,
and hence Eq.~(\ref{eq:probecc}) should not be used in those cases
either. We note that KOI 3138.01 is a prime example of where these
complications can arise. This planet candidate orbits a 0.08~M$_\odot$
star and has $H' = 0.74$, but it could be uninhabitable due
to extreme tidal heating and/or the early super-luminous phase of its
host star.

A recent study of the habitability of planets of low-mass stars
explored the role of synchronous rotation and found that habitable
conditions may exist at orbital distances smaller than in the
traditional HZ calculations \citep{Yang13}. That study found that a
feedback can develop in which clouds form over the sub-stellar point
and increase the albedo up to $\sim 0.6$. This possibility is
implicitly included in our analysis as we allow for albedos larger
than this value, but also reaffirms the need for better constraints on
the distribution of albedos for habitable planets.

In this study we assumed flux limits based on the
runaway greenhouse for planets with Earth-like inventories of
water. Drier planets may remain habitable at much lower or higher 
fluxes. \cite{Abe11} performed 3D models of dry planet atmospheres and
found that a 1~$\mearth$ planet could be habitable for $F <
415$~W/m$^2$ and instellation values greater than $0.58~S_\oplus$. We did not include this possibility for three reasons. First,
they did not provide flux limits at the outer edge. Second, they only considered Earth-radius planets. Third, dry
planets have wider habitable limits and so we expect the peak in $H'$
to be similar for wet planets and dry planets in terms of
$S_{circ}$. Thus, the middle of the ``wet'' HZ of \cite{Kopparapu13}
is close to the middle of the ``dry'' HZ of \cite{Abe11}, and the
comparative habitability of exoplanets using either method should
produce similar rankings. Future work should explore a range of masses
for dry planets and include the limiting flux values.

While we have focused on prioritizing targets for follow-up resources,
our ranking scheme can also prioritize more in-depth theoretical
analyses, such as employing 3-D global climate models
\citep[\eg][]{Pierrehumbert10,Wordsworth11,Shields15}, formation
models \citep[\eg][]{Raymond04,Bond10}, and/or internal evolution
\citep[\eg][]{Behounkova11}. Orbital stability should be determined
through N-body simulations, especially for planets in high
multiplicity systems and/or in mean motion resonances, as the Hill
stability criterion does not apply in those cases and the planets can
evolve chaotically for long periods of time \citep{Barnes15_res}. 

In this study we have explicitly assumed that the potential for
habitability is purely a function of outgoing long-wavelength
radiation ($\sim 1-10 \micron$). This approach is the zeroeth
order model for planetary habitability: Theoretical models of
habitable planets must always reproduce energy balance. A thorough theoretical
exploration could produce a more general ``exoplanetary habitability
index'' that is based on the requirement of liquid surface water for a broad range of planetary compositions and structures, not
emitted fluxes. Future work should develop this concept and apply it
to small planets in and around the classic HZ in order to refine the
priority of exoplanets for biosignature searches by future terrestrial
planet characterization missions, such as {\it
LUVOIR}\footnote{http://science.nasa.gov/media/medialibrary/2013/12/20/secure-Astrophysics\_Roadmap\_2013.pdf}
or the {\it High Definition Space Telescope}
\citep{Dalcanton15}.

The timescale for life to arise is also important in the search for life in the unvierse. In general this timescale is unknown, but can be crudely constrained
by Earth's evolution, which included the unlikely Moon-forming impact
and an orbital instability in the outer Solar System
\citep{Gomes05}. For planets in the HZs of FGK stars, the
interval between the stellar birth and the last giant impact is $\sim
100$~Myr~\citep{KokuboIda98,Raymond04}. These events melted the entire
Earth's surface for up to several Myr
\citep{Zahnle15} and most likely sterilized our planet. Thus, this value is a reasonable choice for the minimum age of a potentially habitable planet, and so, if available, the stellar age should be used as another data point for assessing potential habitability.

In addition to theoretical simplifications, we also do not include any
uncertainties in our analysis. Stellar parameters are notoriously
difficult to constrain, which can often lead to significant
uncertainty in the physical and orbital properties of the planet
\citep{Gaidos13,Kane14}. One should also propagate uncertainties when
calculating $H$, possibly using a Markov chain Monte Carlo approach
that produces quantified posterior distributions
\citep[\eg][]{Kundurthy11}. This investigation is largely conceptual, but including uncertainties would be essential to properly allocate follow-up resources.

Planets with large values of $A$ likely have thick cloud and/or
haze layers that reflect the stellar radiation. These features can
make transit spectroscopy of near-surface atmospheric layers extremely
difficult
\citep[\eg][]{Pont08,Misra14Refraction}. Hence targets that require
large values of $A$ may be poor {\it JWST} targets. Future research
should explore the detectability of atmospheric gases of exoplanets
that require high albedo in order to be habitable. Trace gas
biosignatures that are confined to the troposphere may be below cloud
and/or haze layers are probably undetectable with {\it JWST}.

In rare cases, the acceleration of the planet through transit can be
measured and reveal the true value of $e$
\citep{Barnes07,DawsonJohnson12}. For very high signal-to-noise and
high cadence data, the changes in the durations and slopes of ingress
and egress are detectable. When combined with $\Delta$, these features
break the degeneracy between eccentricity and longitude of pericenter
described in $\S$~\ref{sec:habidx}.2. \cite{Barnes07} estimates that a
Jupiter-mass planet that transits during the acceleration maximum will
require $\sim 3$ ppm precision in the photometry and is thus a
challenging observation to make. Nonetheless for high-priority targets
that are not accessible to RV measurements, such photometric
observations could be very valuable as they could also break the
eccentricity-albedo degeneracy described in $\S$~\ref{sec:habidx}.6.

Finally, we considered known and predicted potentially habitable
exoplanets for {\it JWST} reconnaissance. The best KOI for {\it JWST}
is 5554.01 with $J = 10.1$. This system is probably too faint for
planetary spectral measurements, but more investigations (including
validation) are required to confirm this assessment. Some {\it K2}
targets have already been discovered orbiting brighter stars, but
their locations on the ecliptic equator make any exoplanet in those
fields unlikely to be worthy of {\it JWST} follow-up. Most likely, a
proper target will be discovered by {\it TESS}, and, following up on
\cite{Sullivan15}, we predict that if {\it JWST} obtains transit
transmission spectroscopy of the 2 best planets, then 1 will be potentially
habitable.

\section{Conclusions}
\label{sec:concl}

We have developed a simple metric, the habitability index for
transiting exoplanets, to quantify the likelihood that a transiting
exoplanet may possess liquid surface water. As opposed to the HZ,
which is binary, this approach produces a continuum of values. $H$
can be used to prioritize follow-up efforts, whether they be
observational or theoretical. Our approach to comparative habitability
assessments parameterizes the eccentricity-albedo degeneracy by
calculating the fraction of all possible ($A,e$) pairs that could
produce a clement climate. We must also assess ``rockiness''
probabilistically based on inferences from the few planets with
well-known masses and radii. Despite these assumptions, $H$ can provide more
insight into a planet's potential to support life than simply
comparing its orbit to that of its host star's HZ.

We ranked the known {\it Kepler} and {\it K2} planets for habitability
and found that several have larger values of $H$ than Earth. This does not mean these planets are ``more habitable'' than Earth -- it means that an Earth twin orbiting a solar twin that is observed by {\it Kepler} would not have the highest probability of being habitable. The best
candidates have incident radiation levels, assuming circular orbits,
of 60--90\% that of Earth's. These levels are about in the middle of
the HZ, which is not surprising as our flux limits are derived
from the models that produced the HZ. Our method is grounded in the
fundamental observables and can, when applicable, fold in eccentricity
constraints, and is therefore more powerful than the HZ for transiting
exoplanets.

As the first spacecraft capable of performing transit transmission
spectroscopy is {\it JWST}, we also considered some of its features in
relation to $H$. The detector capabilities and field of regard
constraints could render planets with high values of $H$
unobservable. Thus, decisions on utilization of follow-up resources
should also consider the design of telescopes capable of performing
transmission spectroscopy. We specifically considered 4
candidates for {\it JWST} spectroscopy, KOI 5554.01, KOI
6108.01, KOI 7587.01 and K2-3 d, and find that even
detection of water bands will be challenging for these objects. For
now, the search continues for a suitable target for {\it JWST}.

The characterization of the atmosphere of a rocky exoplanet in the HZ
will mark an important achievement in the history of exoplanetary
science. In the near term, NASA and ESA have developed a sequence of missions
that is capable of achieving this feat, but the resources required to
realize this goal are substantial. The methodology described here can
optimally focus these resources so that we can identify the best
targets for transit transmission spectroscopy as quickly as possible.

\vspace{1cm}
We thank Pramod Gupta for designing the web interface that calculates
the HITE, and Eric Agol, Kevin Zahnle, Rodrigo Luger, Abel M{\'
e}ndez, Ren{\' e} Heller, Drake Deming, Mark Claire and the entire
Virtual Planetary Laboratory for insightful discussions. This research
has made use of the NASA Exoplanet Archive, which is operated by the
California Institute of Technology, under contract with the National
Aeronautics and Space Administration under the Exoplanet Exploration
Program. This work was supported by the NASA Astrobiology Institute's
Virtual Planet Laboratory under Cooperative Agreement No. NNA13AA93A.

\bibliography{CompHab}

\end{document}